\documentclass{article}

\usepackage{arxiv}

\usepackage{url}            
\usepackage{algorithmic}
\usepackage{amsmath}
\usepackage{amsfonts}
\usepackage[small,bf]{caption}
\usepackage[colorinlistoftodos]{todonotes}
\usepackage{hyperref}
\usepackage{float}
\usepackage{url}            

\usepackage{parskip}  

\DeclareFontFamily{U}{wncy}{}
\DeclareFontShape{U}{wncy}{m}{n}{<->wncyr10}{}
\DeclareSymbolFont{mcy}{U}{wncy}{m}{n}
\DeclareMathSymbol{\comb}{\mathord}{mcy}{"58} 

\usepackage{accents}
\newlength{\dhatheight}

\title{Utilizing the Structure of the Curvelet Transform with Compressed Sensing}

\author{
  Nicholas Dwork\thanks{www.nicholasdwork.com, nicholas.dwork@ucsf.edu} \\
  Departments of Biomedical Informatics and Radiology \\
  University of Colorado Anschutz
    \And
  Peder E. Z. Larson \\
  Department of Radiology and Biomedical Imaging \\
  University of California in San Francisco
}

\begin{document}
\maketitle
\setcounter{footnote}{0}

\begin{abstract}
  The discrete curvelet transform decomposes an image into a set of fundamental components that are distinguished by direction and size as well as a low-frequency representation.  The curvelet representation is approximately sparse; thus, it is a useful sparsifying transformation to be used with compressed sensing.  However, the low-frequency portion is seldom sparse.  This manuscript presents a method to modify the redundant sparsifying transformation comprised of the wavelet and curvelet transforms to take advantage of this fact with compressed sensing image reconstruction.  Instead of relying on sparsity for this low-frequency estimate, the Nyquist-Shannon theorem specifies a square region centered on the $0$ frequency to be collected, which is used to generate a blurry estimate.  A Basis Pursuit Denoising problem is solved to determine the missing details after modifying the sparsifying transformation to take advantage of the known fully sampled region.  Improvements in quality are shown on magnetic resonance and optical images.
\end{abstract}



\section{Introduction}
\label{sec:intro}

Fourier sensing applications include Magnetic Resonance Imaging, Optical Coherehence Tomography, Computed Tomography, and Radio Astronomy.  In each of the applications, the data collected are values of the Fourier transform of the image of interest.  To reconstruct the image, one must perform an approximation to the inverse Fourier transform.  It is necessarily an approximation because the samples are a finite set of isolated points in the Fourier domain.  If the samples are collected on a uniform grid, then the inverse Discrete Fourier Transform serves as a fast and high-quality method for image reconstruction.  However, if the samples are not on a uniform grid or if the grid is not fully-sampled (meaning that the samples are not sufficiently dense to satisfy the Nyquist-Shannon sampling theorem), then other inversion algorithms have shown improved results.

Compressed sensing is one such reconstruction method\footnote{Compressed sensing is also known as compressive sampling.}; it permits accurate reconstructions of images with fewer samples than the number required to satisfy the Nyquist-Shannon theorem \cite{adcock2021compressive}.  Compressed sensing can achieve this feat because it incorporates additional knowledge into the reconstruction.  It relies on the \textit{a priori} knowledge that the image can be reconstructed well with a linear transformation applied to a sparse vector.  Common choices include the inverse wavelet transform \cite{lustig2008compressed,baron2018rapid}, the inverse curvelet transform \cite{ma2010improved}, and learned dictionaries \cite{lee2007efficient}.  It is often beneficial to have fast algorithms that implement the sparsifying transformation and fast algorithms exist for the wavelet \cite{mallat1999wavelet,beylkin2009fast} and curvelet transforms \cite{candes2006fast}.  Therefore, we focus on these transforms for this manuscript.  Image reconstruction is formulated as a Basis Pursuit Denoising (BPD) convex optimization problem \cite{candes2011compressed}.  Compressed sensing has benefited many applications including medical imaging \cite{lustig2007sparse}, holography \cite{brady2009compressive}, photography \cite{oike2012cmos}, and communications \cite{huang2013applications,mahalati2013resolution}.

Though the wavelet transforms of natural images are often sparse \cite{majumdar2012choice}, the portion corresponding to the lowest-frequency bin is seldom sparse.  In \cite{dwork2021utilizing}, Dwork et al. used this property with Fourier Sampling to specify the size and shape of a region centered on the $0$ frequency that is fully-sampled, and to alter the optimization problem to solve for a modified vector.  The resulting problem is then converted into the standard form of the BPD problem.  Thus, image reconstruction becomes a two step process:  1) estimate a blurry image with the fully-sampled low-frequency portion of the sampled region, and 2) enhance the image with high frequency details that result from solving a BPD problem.  The system matrix of the optimization problem remains the same, thus all theoretical guarantees of compressed sensing that pertain to the original BPD problem apply to the modified problem.  Since the sparsity of the optimization variable in the new problem is higher than that of the original problem, the error of the result is reduced \cite{adcock2017breaking}.

Redundant dictionaries (also called overcomplete bases) offer a representation that permit a more sparse representation than a basis \cite{candes2011compressed}.  In this work, we consider the overcomplete basis comprised of the inverse wavelet and inverse curvelet transformations for two-dimensional image reconstruction problems.  Notably, like the wavelet transform, although the curvelet transform of a natural image is sparse there is a low-frequency portion of the transform is not sparse.
In this work, we extend the technique of \cite{dwork2021utilizing} and simultaneously take advantage of this low-frequency structure in both transforms\footnote{An early version of this work was presented at the 2021 annual meeting of the Society for Industrial and Applied Mathematicians.}.  This leads to an incremental improvement in quality at the cost of additional computations required for the curvelet transform and its inverse.  Again, the theoretical guarantees of compressed sensing that pertain to reconstruction with a redundant vector set remain.  The increased sparsity attainable with the redundant dictionary further reduces the theoretical bound on the error.  We will demonstrate a reduced error empirically with several images.

\section{Methods}
\label{sec:methods}

In this section, we review the basis pursuit denoising optimization problem and the theoretical results of compressed sensing.  We will then discuss compressed sensing using the redundant dictionary comprised of wavelet and curvelet vectors.  And we will show a method to utilize the structure of the sparsifying transform to better sample the Fourier domain and increase the sparsity of the optimization variable.  Results of reconstructing images with these algorithms are presented in section \ref{sec:results}.

\subsection{Background}
\label{sec:background}

Let $x$ be the desired image and $b$ be the measured data where $A\,x + \epsilon = b$.  Here, $A$ is called the system matrix and $\epsilon$ is an additive Gaussian noise vector of independent and identically distributed values.  Let $\Psi$ be the sparsifying linear transformation.   Let $y$ be a sparse vector such that the desired image is approximated well by $\Psi^\ast y$.  Then one can estimate $y$ by solving the following optimization problem:
\begin{equation}
  \underset{y}{\text{minimize}} \hspace{0.5em} \|y\|_0 \hspace{0.5em}
    \text{subject to} \hspace{0.5em} (1/2)\|A \, y - b\|_2 \leq \sigma,
  \label{prob:sparseSignalRecovery}
\end{equation}
where $\sigma$ is a bound on the noise power.  With this notation, $\Psi^\ast$ is included in $A$ (see below).  This problem, called the sparse signal recovery problem, is an np-hard combinatorial optimization problem.  The BPD problem is a related problem where the $L_0$ penalty is replaced with the $L_1$ norm:
\begin{equation}
  \underset{y}{\text{minimize}} \hspace{0.5em} \|y\|_1 \hspace{0.5em}
    \text{subject to} \hspace{0.5em} (1/2)\|A y - b\|_2 \leq \sigma.
  \label{prob:bpd}
\end{equation}
This is a convex optimization problem and can be solved with known efficient algorithms.

One might hope that solving \eqref{prob:bpd} gets us close to an answer for \eqref{prob:sparseSignalRecovery}.  Indeed, the theory of compressed sensing dictates that when $A$ satisfies specific properties (e.g. the Mutual Coherence Conditions \cite{donoho2003optimally}, the Restricted Isometry Property \cite{candes2008introduction}, or the Restricted Isometry Property in Levels \cite{adcock2017breaking}) then the solution to the BPD problem \textit{is optimal} for the sparse signal recovery problem with very high probability.  Moreover, the more sparse $y$ is, the fewer the number of measurements that are required to attain the solution \cite{candes2008introduction,adcock2017breaking}.  This latter fact is the leverage that we will exploit to generate high quality images with fewer samples in this manuscript.

\begin{figure}[ht]
  \centering{}
  \includegraphics[width=0.8\linewidth]{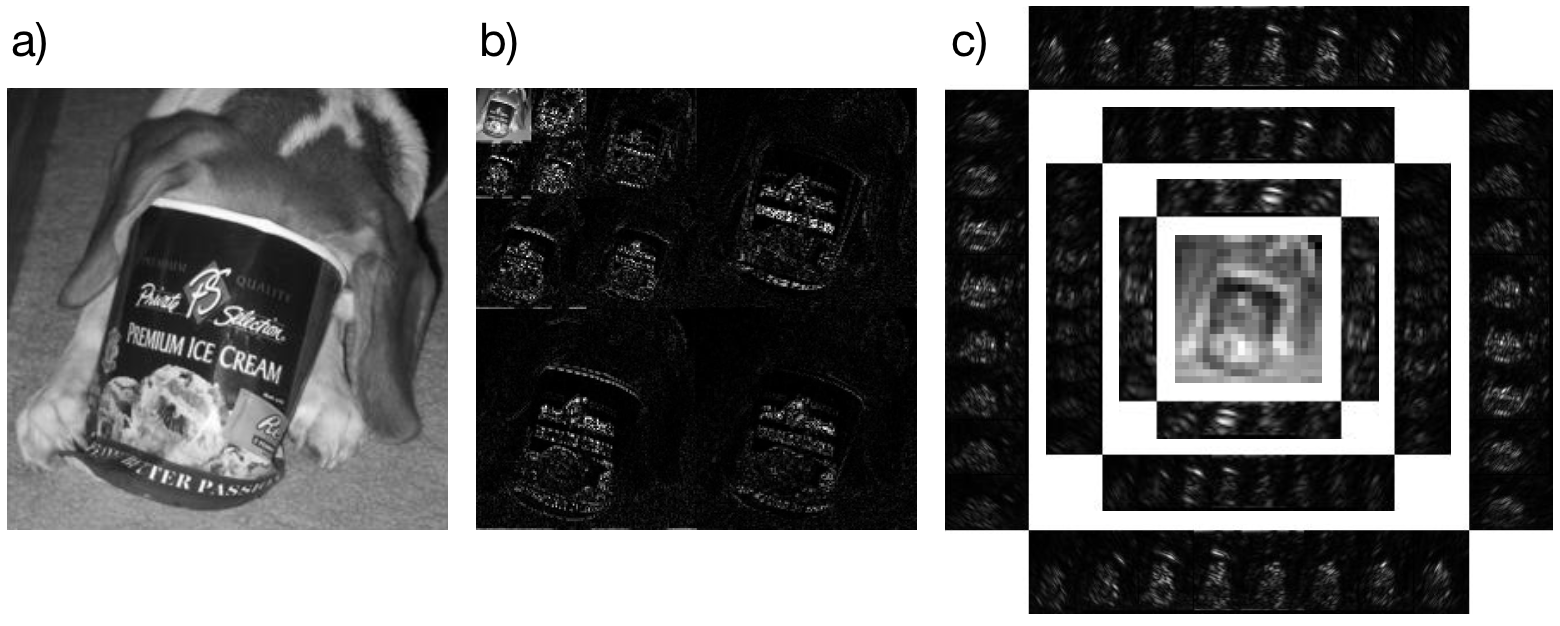}
  \captionsetup{width=.95\linewidth}
  \caption{ \label{fig:bacWavCurv} (a) Original image, (b) Magnitude of the discrete Daubechies-4 wavelet transform coefficients, (c) Magnitude of the curvelet transform coefficients. }
\end{figure}

With standard compressed sensing and Fourier sensing, $b$ is the vector of measured Fourier values and $A = M\, F\, \Psi^\ast$, where $\Psi$ is the sparsifying transformation (which may have more rows than columns), $\cdot^\ast$ represents the adjoint of the transformation, $F$ is the unitary Discrete Fourier Transform and $M$ is the sampling mask that identifies those samples that were collected.  An effective choice for $\Psi$ is the Discrete Daubechies Wavelet Transform (DDWT).  Figure \ref{fig:bacWavCurv}a shows an image of a dog eating from an ice cream container, and Fig. \ref{fig:bacWavCurv}b shows the magnitude of the DDWT of order $4$ (DDWT-$4$) coefficients when applied recursively to the lowest-frequency bin $4$ times.  The vast majority of the resulting coefficients appear black, which demonstrates the sparsifying behavior of the DDWT-4.  The upper left corner of the transform is not black; this is a low-pass filter and downsampling of the original image, and is seldom sparse with natural images.

Let $W$ represent the orthogonal DDWT-$4$.  To satisfy the Nyquist-Shannon sampling theorem in order to accurately estimate the lowest frequency bin of the DDWT-$4$, one must collect a fully-sampled region with size equal to that of the lowest-frequency bin \cite{dwork2021utilizing}.  This is the same fully-sampled region as specified by the two-level sampling scheme in \cite{adcock2017breaking}.  From this fully-sampled region, a blurry image can be reconstructed \cite{dwork2021utilizing}.  When using the DDWT-$4$ as the sparsifying transformation, optimizing for the remaining details increases the sparsity of the resulting optimization variable.  The missing details of the image can be estimated by solving the following problem:
\begin{equation}
  \underset{y}{\text{minimize}} \hspace{0.5em} \| y \|_1 \hspace{0.5em}
    \text{subject to} \hspace{0.5em} \|M\,F\,\left( \Psi^\ast y + x_L \right) - b\|_2 \leq \sigma,
  \label{prob:msbpd}
\end{equation}
where $\Psi=W$, and $x_L = F^\ast \, K_B \, M_L \, b$.  Here, $K_B$ is a Kaiser-Bessel window used to reduce ringing in the blurry estimate and $M_L$ is a mask that specifies the fully-sampled low-frequency region.  Problem \eqref{prob:msbpd} is equivalent to
\begin{equation}
  \underset{y}{\text{minimize}} \hspace{0.5em} \|y\|_1 \hspace{0.5em}
    \text{subject to} \hspace{0.5em} (1/2)\|M \, F\, \Psi^\ast\, y - \beta\|_2 \leq \sigma,
  \label{prob:msbpdAsBPD}
\end{equation}
where $\beta = b - M \, F \, x_L$.  By letting $A=M\,F\,\Psi^\ast$, it becomes apparent that this problem has the form of the BPD problem \eqref{prob:bpd} where $b$ has been replaced by $\beta$.
Once the solution $y^\star$ is determined, the completely reconstructed image is $x^\star = x_L + \Psi^\ast \, y^\star$.  We call problem \eqref{prob:msbpdAsBPD} the Structured BPD (S-BPD) problem.

\subsection{Curvelets}

Curvelets offer an opportunity to further increase the sparsity of the optimization variable.  The curvelet transformation permits a representation of an image using a multi-scale pyramid with several directions available at each position (as opposed to wavelets, which decompose an image into horizontal and vertical components).  Note that the discrete curvelet transform is a tall transform (it is a tight frame \cite{adcock2019frames}) with more coefficients than there are pixels in the original image.  Using curvelets as the sparsifying transform with compressed sensing image reconstruction has been shown to reduce the blocking artifacts common to wavelet representations \cite{liew2004blocking}.
Figure \ref{fig:bacWavCurv}c shows the magnitude of the discrete curvelet coefficients of Fig. \ref{fig:bacWavCurv}a.  (Note that the display of the curvelet coefficients includes white-space to ease interpretation; it was generated using the \verb|CurveLab| software \cite{curvelab}.)

Similar to the low-frequency bin of the wavelet transform, the center portion of the curvelet transform is the result of windowing, low-pass filtering, and downsampling the original image \cite{candes2006fast}.  Thus, like the wavelet transform, this region will not be sparse for most natural images.

\subsection{Structured Sparsity with Curvelets and Wavelets}
\label{subsec:curveletSparsifier}

Consider the redundant dictionary comprised of curvelets and wavelets.  The number of significant non-zero coefficients would never increase with the addition of curvelets into the dictionary for the image can always be represented with the same wavelet coefficients and with all curvelet coefficients equal to $0$.
Intuitively, one would expect that increasing the redundancy of the set of vectors would permit an even more sparse representation of an image.  Suppose, for example, one wanted to represent $v = w_i + c_j$ where $w_i$ is an element of the wavelet basis and $c_i$ is an element of the curvelet basis \cite{starck2002curvelet}.  If one were to express $v$ as a linear combination of vectors from both bases then the number of non-zero linear coefficients required would be $2$.  However, if one were to try to represent the same image $v$ using only the wavelet basis, then it would require a larger number of significant linear coefficients.  

Recall that, in accordance with the theorems of compressed sensing, increased sparsity permits an accurate reconstruction with fewer data samples.  Indeed, after the benefit was shown empirically \cite{peyre2010best,doneva2010compressed}, the theory was developed to show that accurate reconstruction was possible with a redundant (or overcomplete) incoherent set of vectors \cite{candes2011compressed,rauhut2008compressed}.  We will now present a method analogous to that of \cite{dwork2021utilizing} that simultaneously takes advantage of the low-frequency structure present in both the curvelet and wavelet transforms of natural images.

\begin{figure}[ht]
  \centering{}
  \includegraphics[width=0.8\linewidth]{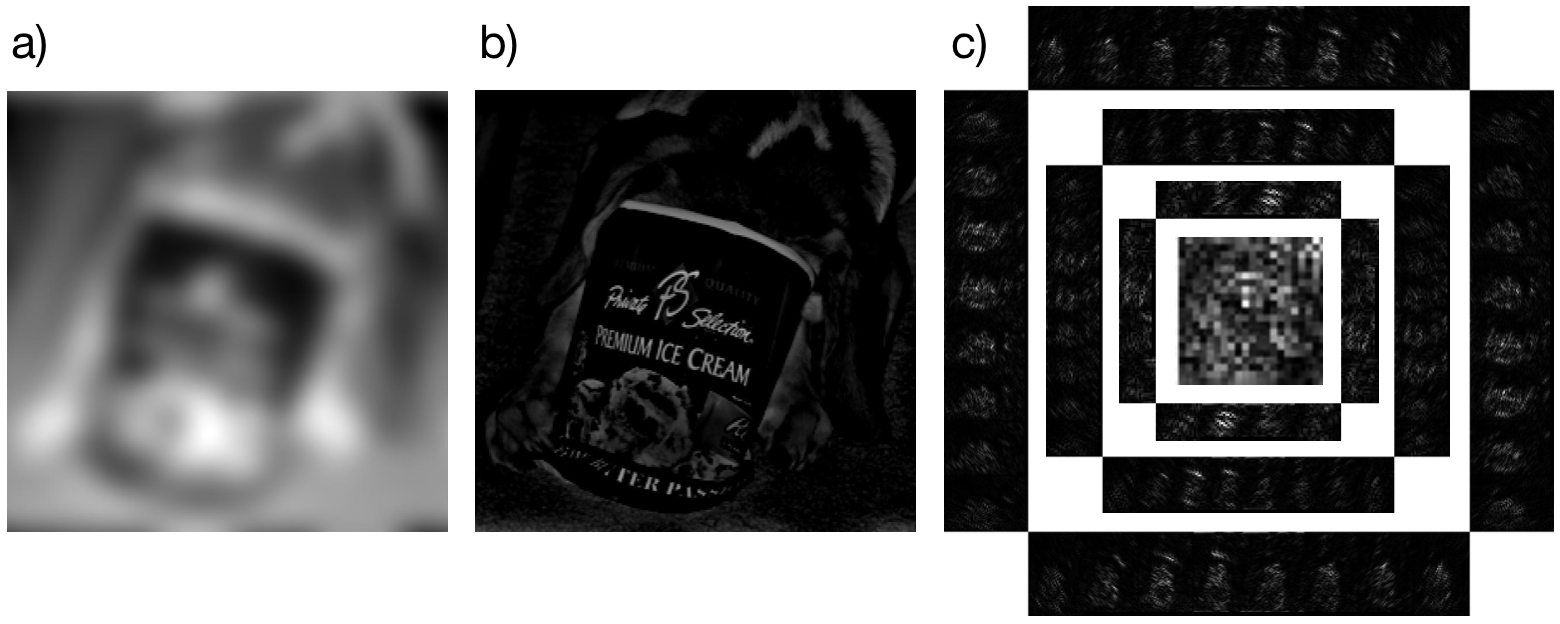}
  \captionsetup{width=.95\linewidth}
  \caption{ \label{fig:cSparserBac} (a) Blurry estimate, (b) Result of subtracting the blurry estimate from the original image, (c) Magnitude of the curvelet transform coefficients of b.  Note that the low-frequency center region in c is much more sparse than the corresponding region of Fig. \ref{fig:bacWavCurv}c. }
\end{figure}

Since sparsity cannot be assumed for low-frequency region of the Curvelet transform, we rely on the Nyquist-Shannon sampling theorem.  This dictates a fully-sampled region with spacing equal to the inverse of the full size of the image and with size equal to the center portion of the curvelet transform.  The values of the fully-sampled region are used to create a blurry image by applying a low pass window ($W_0$ as specified in \cite{candes2006fast}) and then performing an inverse discrete Fourier Transform.  Figure \ref{fig:cSparserBac}a shows this blurry estimate for the image of Fig. \ref{fig:bacWavCurv}a.  Figure \ref{fig:cSparserBac}b shows the result of subtracting this blurry estimate from the original image, and Fig. \ref{fig:cSparserBac}c shows the curvelet transform of Fig. \ref{fig:cSparserBac}b.  When comparing Fig. \ref{fig:bacWavCurv}c with Fig. \ref{fig:cSparserBac}c, the increased sparsity of the low-frequency region is apparent.

This is usually the case for natural images: after subtracting the blurry estimate from the original image, the curvelet transform of the result  sparsity of the curvelet transform is increased.

As discussed, both wavelets and curvelets have a region that is a low-pass filter and downsampling of the original image.  The larger of these two regions dictates a fully-sampled region centered on $0$ frequency to be collected that satisfies the Nyquist-Shannon sampling theorem.  From this fully-sampled region, a blurry frequency estimate is created.  If the center of the curvelet is larger than the lowest-frequency bin of the wavelet transform, then the blurry estimate is generated according to subsection \ref{subsec:curveletSparsifier}.  Otherwise, the blurry estimate is generated according to \cite{dwork2021utilizing}.  Denote this blurry estimate as $x_L$.
Let $\Psi=(W,C)$, the block-matrix comprised of the wavelet and curvelet transforms.  As discussed, the columns of $\Psi^\ast$ may offer a more sparse representation of the image.  This redundancy makes a more sparse representation of the optimization variable available, leading to a reconstruction with a reduced error bound.  The image is reconstructed by solving \eqref{prob:msbpdAsBPD} for $y^\star$ and setting $x^\star = x_L + \Psi^\ast \, y^\star$.

\section{Experiment Design}
\label{sec:experiments}

\begin{figure}[ht]
    \centering
    \includegraphics[width=0.95\linewidth]{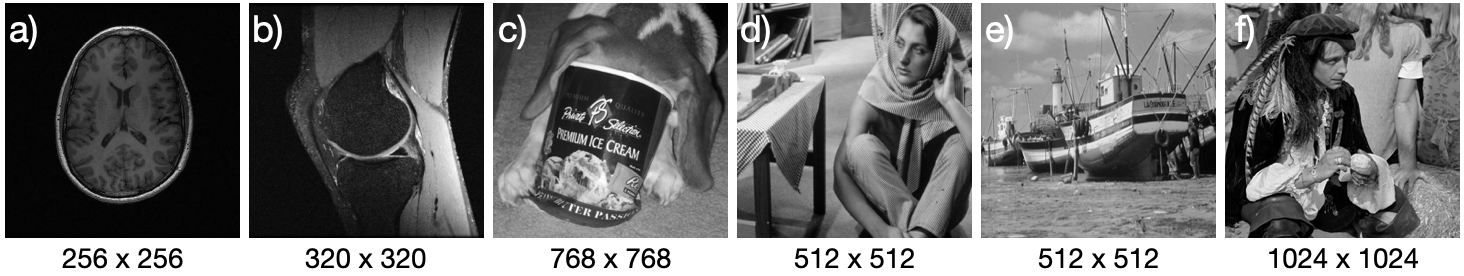}
    \vspace{0.25em}
    \caption{The six images analyzed in this manuscript along with their sizes in pixels.  The data of subfigures (a) and (b) were collected with MRI machines; the data of subfigures (c)-(f) were collected with optical cameras.  The data of subfigure (b) was made available to the public at \url{mridata.org}.}
    \label{fig:imgsAnalyzed}
\end{figure}

The six images studied in this manuscript, along with their sizes, are shown in Fig. \ref{fig:imgsAnalyzed}.
All images were scaled by their maximum so that pixel values lied within $[0,1]$.  The data for Fig. (a) and (b) were collected with Magnetic Resonance Imaging (MRI) machines.  The data for Fig. \ref{fig:imgsAnalyzed}a were collected with a $3$ Tesla clinical MRI using a Cartesian trajectory, a field of view of $25.6\times 25.6$ cm$^2$, and a $1$ mm slice thickness.  The data of Fig. \ref{fig:imgsAnalyzed}a were gathered with Institutional Review Board (IRB) approval, Health Insurance Portability and Accountability Act (HIPAA) compliance, and patient informed assent/consent.  The data for Fig. \ref{fig:imgsAnalyzed}b was made available to the public at \url{mridata.org} \cite{ong2018mridata}. The data for subfigures \ref{fig:imgsAnalyzed}c-f were made publicly available from various sources and are often studied in image processing research papers.  All data was initially fully-sampled and retrospectively downsampled for processing with the algorithms presented in this manuscript.
The MRI data of the knee (Fig. \ref{fig:imgsAnalyzed}b) is from a fully-sampled three-dimensional dataset.  The data was transformed into a the hybrid-space of spatial frequency in the anterior-posterior and superior-inferior dimensions and location in the left-right dimension.  Afterwards, a two-dimensional slice located near the center of the knee in the left-right dimension was isolated for further processing in this manuscript.

Other works have attempted to address the difference in sparsity of the lowest-frequency bin of the sparsifying transform by minimizing an objective function with a weighted norm, as follows:
\begin{equation}
  \underset{y}{\text{minimize}} \hspace{0.5em} \|y\|_{\omega,1} \hspace{0.5em}
    \text{subject to} \hspace{0.5em} (1/2)\|A y - \beta\|_2 \leq \sigma,
  \label{prob:weightedBPD}
\end{equation}
where $\|y\|_{\omega,1}=\omega_1 |y_1| + \omega_2 |y_2| + ... + \omega_N |y_N|$.
In \cite{candes2008enhancing}, Candes et al. iteratively update the weight vector $\omega$ by setting it inversely proportional to the intensity of the corresponding coefficient.  In \cite{varela2021automatic}, Varela et al. set the weight $\omega$ for each bin of the Wavelet transform independently by analyzing the zero-filled reconstruction, and they set the elements of $\omega$ for the lowest-frequency bin to $0$.  We will compare to a non-iterative version of these techniques where $\omega_i=0$ if the pixel in the transformed domain corresponds to the lowest-frequency bin and $\omega_i=1$ otherwise.  That is, $\omega$ is a mask that specifies whether or not the pixel is an element of a high-frequency bin of the sparsifying transform.  We will denote solving problem \eqref{prob:weightedBPD} with this definition of $\omega$ as \textit{BPD with Mask}.

For all results presented, the Fast Iterative Shrinkage Threshold Algorithm (FISTA) with line search \cite{beck2009fast,scheinberg2014fast} run for $100$ iterations was used to solve the equivalent Lagrangian form of the problem.  For example, to solve problem \eqref{prob:bpd}, the following equivalent optimization problem was solved:
\begin{equation}
  \text{minimize} \hspace{0.5em} (1/2) \|A\, y - \beta \|_2^2 + \|y\|_{\lambda,1},
  \label{eq:msbpdLASSO}
\end{equation}
where $\lambda$ is the regularization vector.  The iterative re-weighting scheme of \cite{candes2008enhancing} was used to automatically determine $\lambda$ with the number of re-weighting iterations equal to $5$.  The initial value of $\lambda$ was set to the corresponding values of the wavelet coefficients determined with the zero-filled reconstruction (the reconstruction assuming all values not sampled are equal to $0$).
The error metrics reported are relative error, defined as $e = \| \text{true} - \text{estimate} \|_2 / \| \text{true} \|_2$, mean squared error, and the structural similarity metric (SSIM) \cite{wang2003multiscale}.

\begin{figure}[ht]
  \centering{}
  \includegraphics[width=0.75\linewidth]{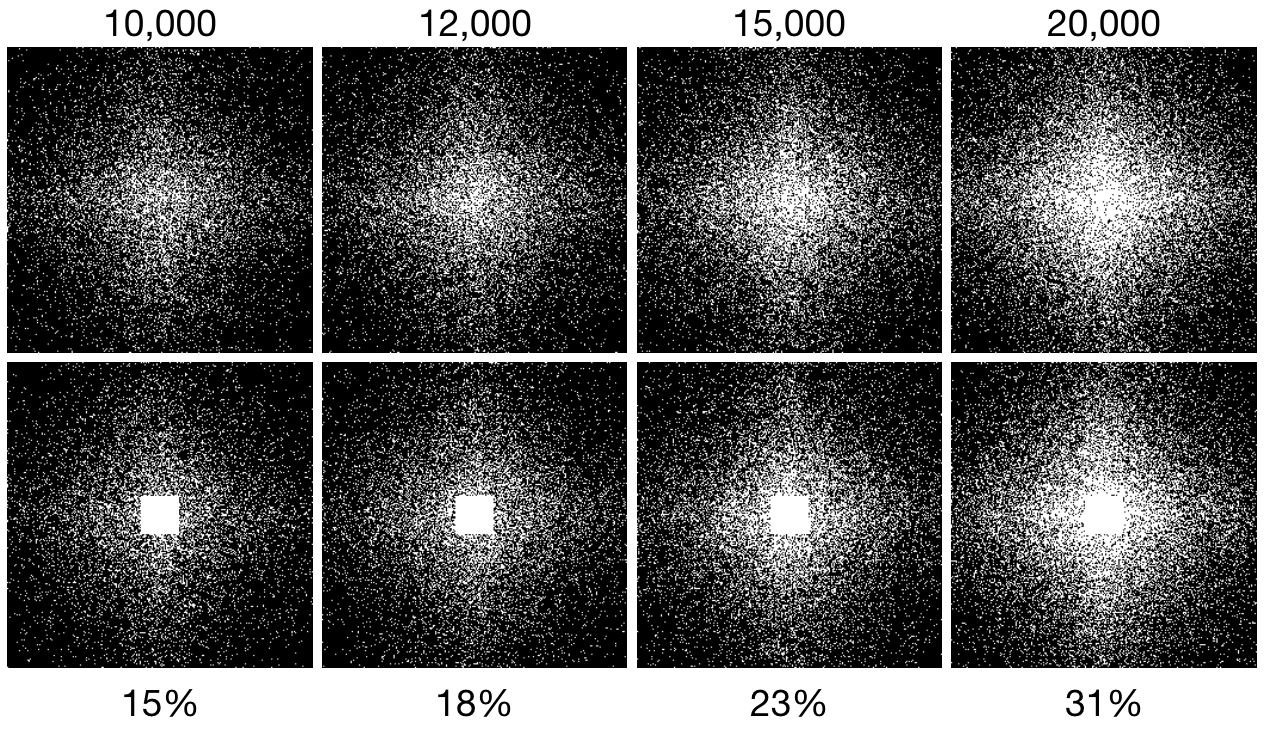}
  \captionsetup{width=.9\linewidth}
  \caption{ \label{fig:samplingPatterns} Example variable density sampling patterns generated from a separable Laplacian distribution with a standard deviation of $0.3$ the length of the sides.  The top/bottom row shows sampling patterns without/with the fully-sampled center region corresponding to the size of the low-frequency region of the curvelet transform, respectively.  From left to right, the sampling patterns include $10,000$, $12,000$, $15,000$, and $20,000$ point, respectively; this corresponds to $15\%$, $18\%$, $23\%$, and $31\%$ of the total number of samples. }
\end{figure}

Three different sparsifying transformations were tested:  DDWT-4, the wrapping version of the discrete curvelet transform, and a redundant dictionary comprised of both of these transforms; we denote these transformations as wavelet, curvelet, and wavCurv, respectively.
For each transformation, two different sampling patters were tested:  a variable density sampling pattern and a variable density sampling pattern with a fully-sampled center region\footnote{Due to spatial limitations, not all results are presented.}.  The sampling patterns used for the results in this manuscript were generated using a separable Laplacian distribution.  Examples are shown in Fig. \ref{fig:samplingPatterns}.  When a fully-sampled region was included, the number of variable density samples was reduced to retain the same total number of samples.

\section{Results}
\label{sec:results}

\begin{figure}[ht]
  \centering{}
  \includegraphics[width=0.8\linewidth]{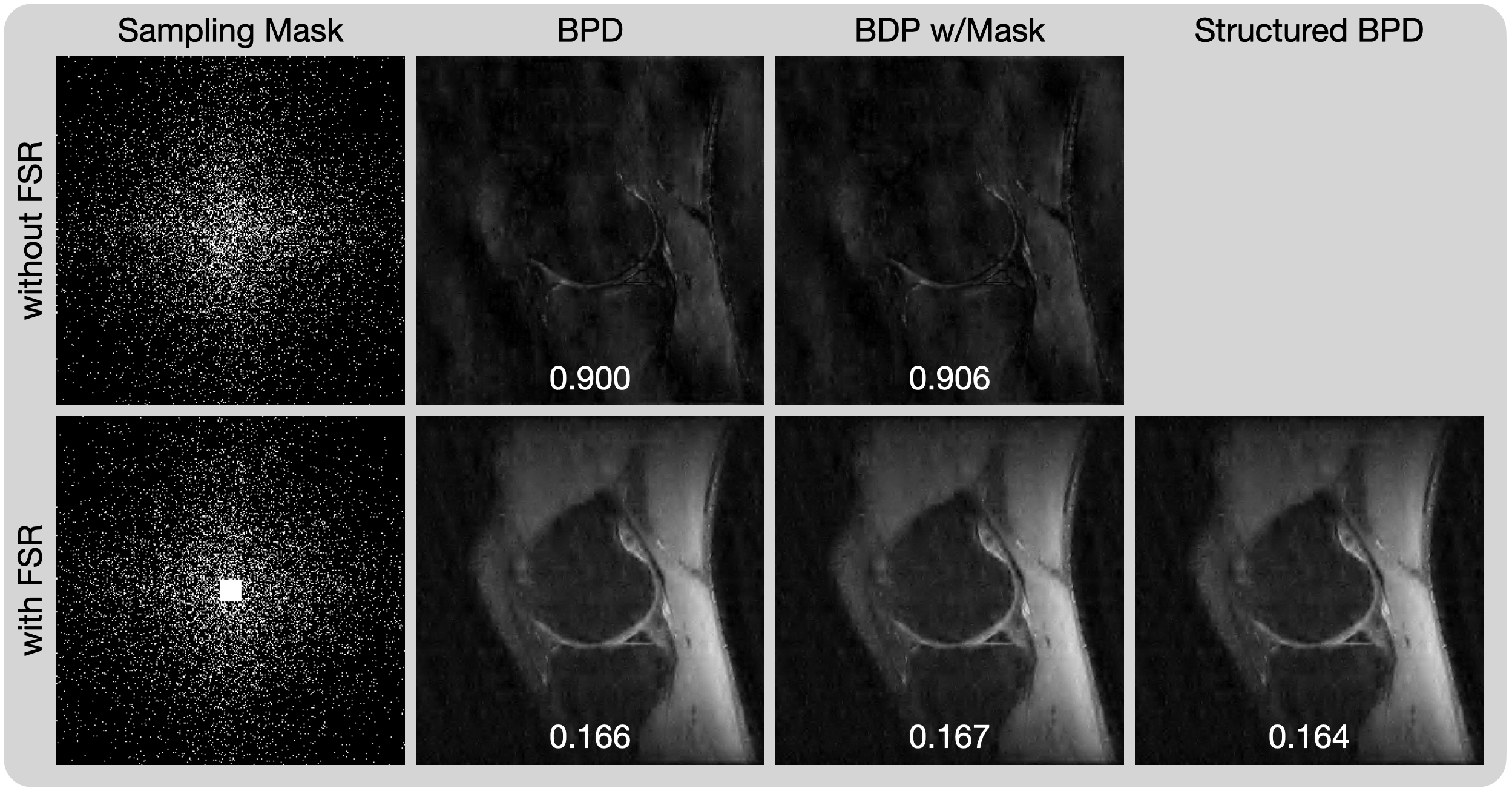}
  \captionsetup{width=.95\linewidth}
  \vspace{0.75em}
  \caption{ \label{fig:kneeRecons} (Left) Sampling masks without and with the fully-sampled center region.  (Right) Reconstructions of the knee with 8192 samples ($8\%$ of the total number of pixels.  The relative error for each reconstruction is written in the corresponding image.}
\end{figure}

Figure \ref{fig:kneeRecons} shows reconstructions of Fig. \ref{fig:imgsAnalyzed}b from $8192$ samples ($8\%$ of the number required for full sampling).  The sampling mask was generated from a separable Laplacian distribution with a standard deviation of $0.3$ the size of the image in each dimension.  The reconstructions from sampling patterns that include the fully-sampled center region (FSR) are far superior than those without it.  When the FSR is included, the relative error for the reconstruction from the Structured BPD problem is slightly smaller than those reconstructed with BPD or BPD with a regularization mask.

\begin{figure}[ht]
  \centering{}
  \includegraphics[width=0.65\linewidth]{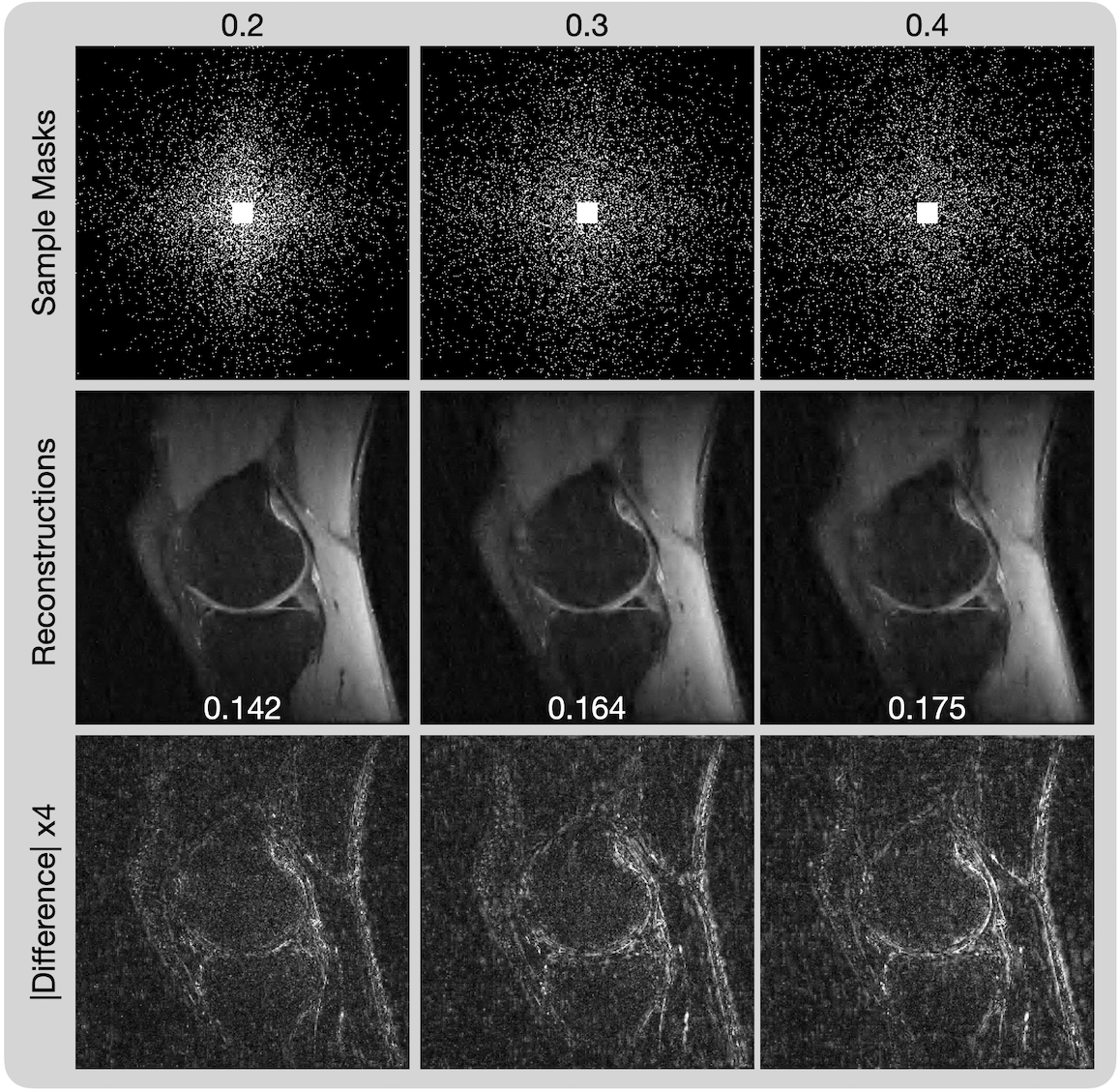}
  \captionsetup{width=.95\linewidth}
  \vspace{0.75em}
  \caption{ \label{fig:kneeReconsVdSigs} Reconstructions of the knee with 8192 sampling patterns generated from a separable Laplacian distribution with a standard deviation of (left) $0.2$, (center) $0.3$, and (right) $0.4$.  The relative error for each reconstruction is written in the corresponding image. }
\end{figure}

Figure \ref{fig:kneeReconsVdSigs} shows reconstructions of Fig. \ref{fig:imgsAnalyzed}b from $8192$ samples with sampling masks generated from separable Laplacian distributions with standard deviations of $0.2$, $0.3$, and $0.4$ along with the magnitude of the difference images.  The standard deviation has a significant effect on the quality of the reconstruction and the metric value.

\begin{figure}[ht]
  \centering{}
  \includegraphics[width=0.60\linewidth]{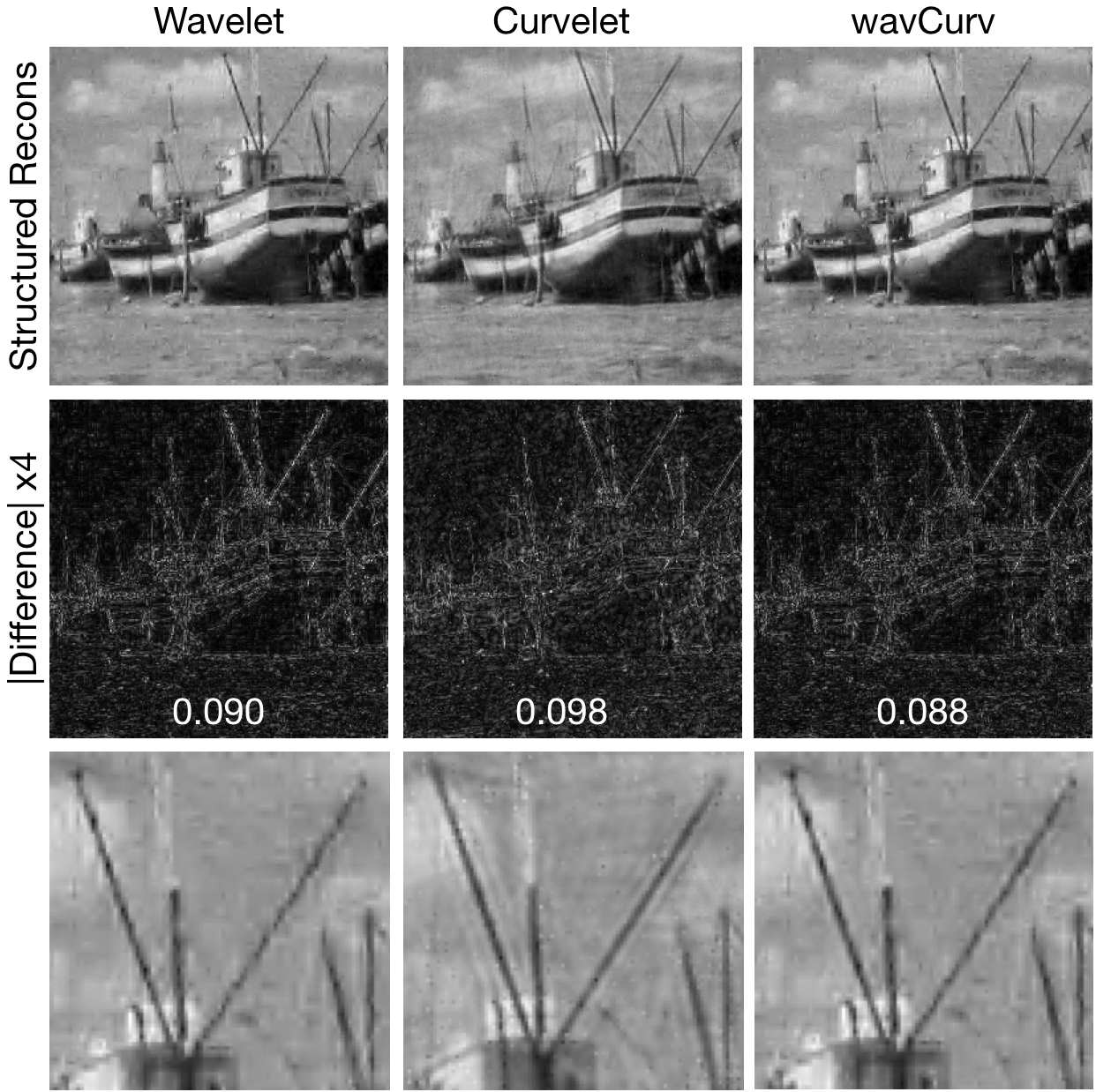}
  \captionsetup{width=.8\linewidth}
  \caption{ \label{fig:boatReconsMSBPD} Reconstructions of Fig. \ref{fig:imgsAnalyzed}e using the structured compressed sensing algorithm using three different sparsifying transformations: wavelets, curvelets, and wavelets and curvelets.  A autocalibration region was included as shown in the second row of Fig. \ref{fig:samplingPatterns} and the Laplacian variable density sampling pattern was generated with a standard deviation of $75$ pixels. }
\end{figure}

Figure \ref{fig:boatReconsMSBPD} shows the results of solving the Lagrangian form of problem \eqref{prob:msbpdAsBPD} for three different sparsifying transformations: wavelets, curvelets, and the redundant dictionary made of both wavelets and curvelets.  We see that the wavelet sparsifying transformation achieves a result with less relative error than the curvelet sparsifying transformation.  Notably, though, the masts of the boat appear sharper when using curvelets than when using wavelets.  However, there are striations in the clouds in the curvelet reconstruction that are not present in the original image or in the wavelet reconstruction.  The lowest relative error is attained by using both wavelets and curvelets, and the reconstruction is able to reduce the block artifacts on the boat mast associated with wavelets without adding striations into the clouds.  The reconstruction from wavelets alone presents blocking artifacts, and the reconstruction from curvelets alone presents striping artifacts with isolated high level values; both artifacts are well known \cite{wieczorek2015x}.  The reconstruction of the redundant dictionary presents a combination of both artifacts, and does so in a way that recapitulates the details of the reference image better.

\begin{figure}[ht]
  \centering{}
  \includegraphics[width=0.8\linewidth]{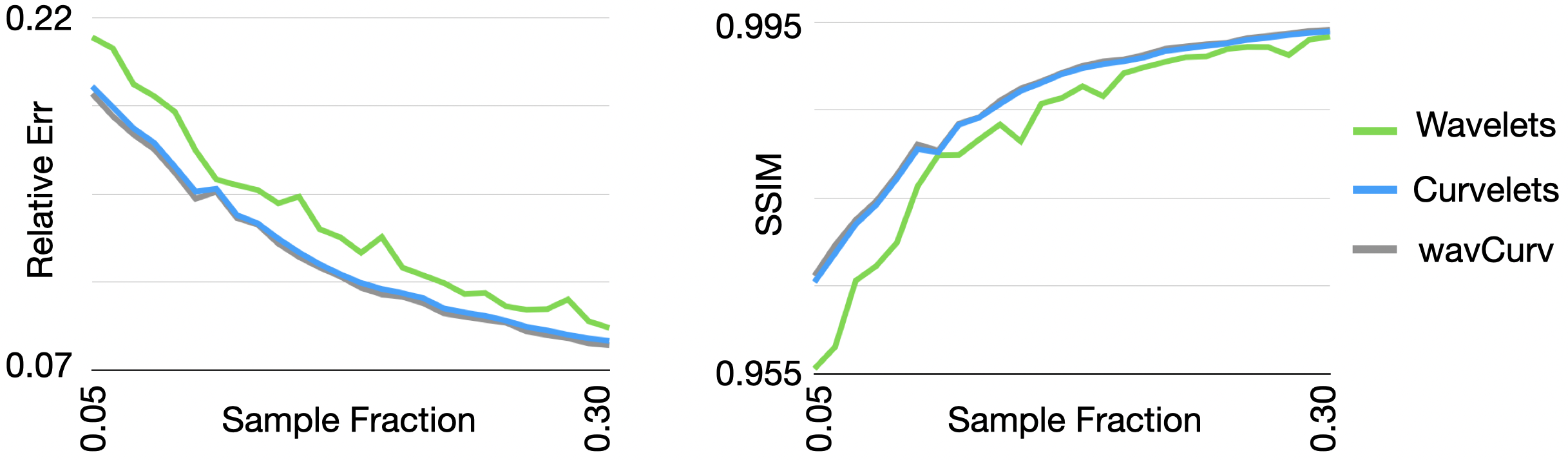}
  \captionsetup{width=.95\linewidth}
  \vspace{1.0em}
  \caption{ \label{fig:brainErrPlots} Plots of error metrics versus sampling percentage for Brain reconstructions }
\end{figure}

Figure \ref{fig:brainErrPlots} shows plots of mean squared error and structural similarity metric for reconstructions of Fig. \ref{fig:imgsAnalyzed}a generated with Structured BPD using sparsifying transformations of wavelets, curvelets, and the redundant dictionary versus number of samples.  The redundant dictionary generates reconstructions with lower (better) relative err and higher (better) structural similarity for all numbers of samples.

Table \ref{tbl:qualityMetrics} shows the mean squared error and structural similarity for reconstructions generated with BPD, BPD with a regularization mask, and Structured BPD using sampling masks without and with the fully sampled center region and $10\%$ of the samples required to satisfy the Nyquist-Shannon sampling theorem.  For all the images of Fig. \ref{fig:imgsAnalyzed}, Structured BPD attains the best quality metrics.

\begin{table}
  \scriptsize
  \begin{center}
  \begin{tabular}{c|c|c|c|c|c|c|c|c|c|c|}
    \cline{2-11}
    & \multicolumn{4}{|c|}{Without FSR} & \multicolumn{6}{|c|}{With FSR} \\
    \cline{2-11}
    & \multicolumn{2}{c}{SSIM} & \multicolumn{2}{|c|}{MSE $\cdot 10^3$} & \multicolumn{3}{c}{SSIM} & \multicolumn{3}{|c|}{MSE $\cdot 10^3$} \\
    \hline
    \multicolumn{1}{|c|}{Image} & BPD & BPD w/Mask & BPD & BPD w/Mask & BPD & BPD w/Mask & S-BPD & BPD & BPD w/Mask & S-BPD \\
    \hline
    \multicolumn{1}{|c|}{a} & 0.687 & 0.613 & 6.20 & 7.20 & 0.968 & 0.967 & 0.969 & 0.656 & 0.670 & 0.638 \\
    \multicolumn{1}{|c|}{b} & 0.876 & 0.871 & 7.88 & 8.18 & 0.981 & 0.981 & 0.981 & 1.24 & 1.25 & 1.22 \\
    \multicolumn{1}{|c|}{c} & 0.790 & 0.788 & 19.3 & 19.4 & 0.973 & 0.972 & 0.973 & 2.95 & 3.05 & 2.92 \\
    \multicolumn{1}{|c|}{d} & 0.00 & 0.00 & 236 & 236 & 0.939 & 0.939 & 0.940 & 6.03 & 6.02 & 6.00 \\
    \multicolumn{1}{|c|}{e} & 0.823 & 0.821 & 12.2 & 12.2 & 0.947 & 0.946 & 0.948 & 4.08 & 4.12 & 4.03 \\
    \multicolumn{1}{|c|}{f} & 0.700 & 0.700 & 26.4 & 26.5 & 0.969 & 0.969 & 0.969 & 3.63 & 3.68 & 3.62 \\
    \hline
  \end{tabular}
  \end{center}
  \vspace{0.5em}
  \caption{Structural Similarity and Mean Squared Error metric values for reconstructions from $10\%$ of samples using a sampling mask generated from a separable Laplacian distribution with a standard deviation of $0.3$ excluding and including an auto-calibration region using a sparsifying transformation comprised of wavelets and curvelets.  The image labels are according to Fig. \ref{fig:imgsAnalyzed}.  The numbers written in blue highlight the best result for the image. }
  \label{tbl:qualityMetrics}
\end{table}

\begin{figure}[ht]
  \centering{}
  \includegraphics[width=0.5\linewidth]{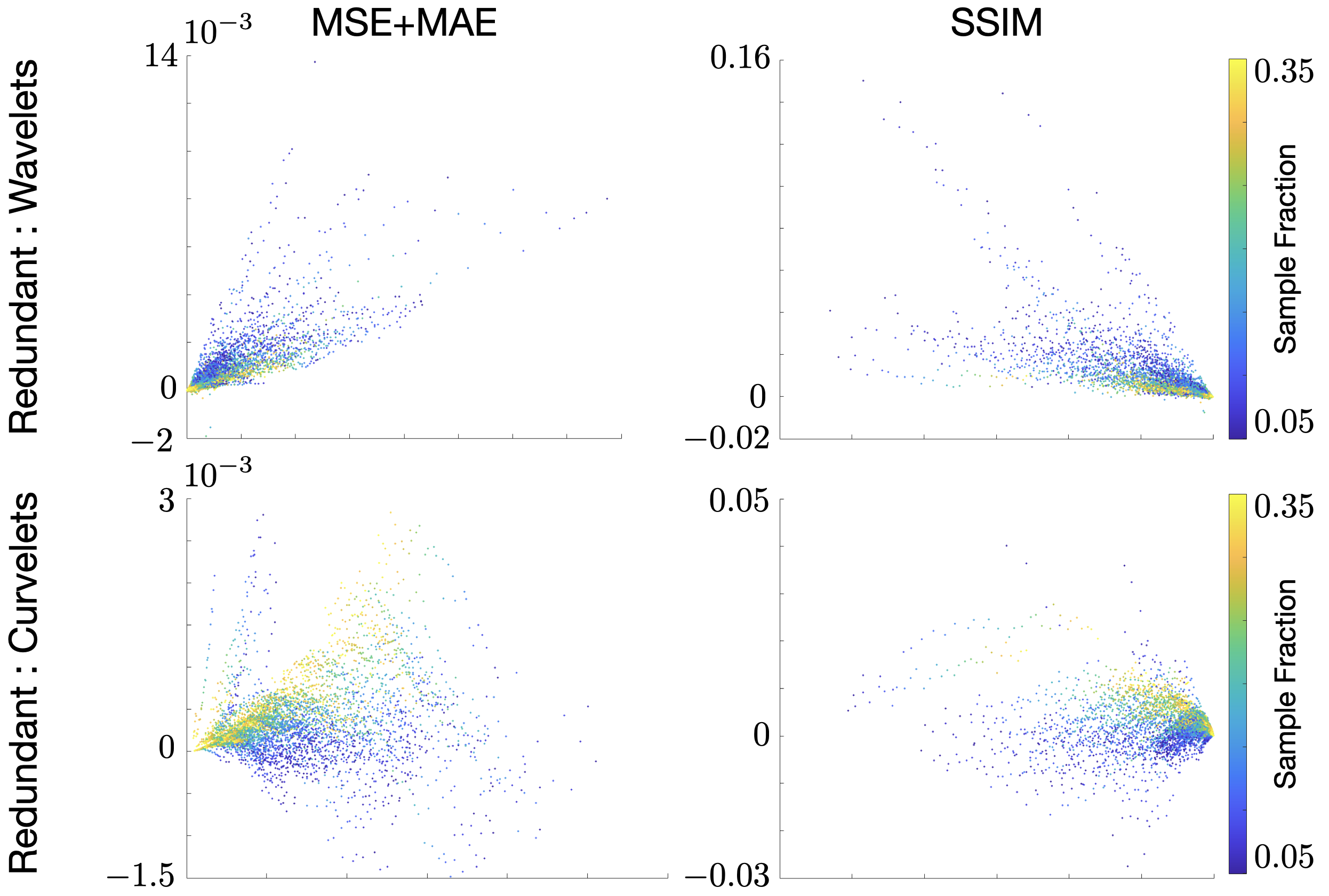}
  \captionsetup{width=.95\linewidth}
  \vspace{1.0em}
  \caption{ \label{fig:improvementMetrics} Improvements in the SSIM and Mean Absolute Error plus Mean Squared Error for images for Structured BPD with the redundant dictionary of wavelets and curvelets plotted against wavelets and curvelets for 158 different images (each from a separate class) taken from the ImageNet dataset \cite{russakovsky2015imagenet}.  The color indicates the fraction of Fourier domain pixels that were sampled. }
\end{figure}

Figure \ref{fig:improvementMetrics} presents differences in the values of SSIM and the sum of mean absolute error (MAE) with mean squared error (MSE) for reconstructions from solving \eqref{eq:msbpdLASSO} when using the redundant dictionary of wavelets and curvelets plotted against a dictionary comprised exclusively of wavelets and a dictionary comprised exclusively of curvelets.  SSIM is a perceptually motivated metric while MAE+MSE attempts to present a more balanced metric than either MAE or MSE alone \cite{wang2022stochastic}.  The results are plotted for 158 different images from the ImageNet database where each image was randomly selected from a separate class \cite{russakovsky2015imagenet}.  The Fourier values of each square image were sampled with an FSR added to a variable density Poisson distribution using a parameter value of 0.3 the length of the image's side.  As can be seen, the SSIM value when using the redundant dictionary is almost always higher than when using wavelets alone, indicating that the quality is improved when using the redundant dictionary.

For the majority of images, the redundant dictionary improves the quality of the result with respect to either of the metrics evaluated.  When comparing the redundant dictionary to wavelets, the improvement is usually more pronounced when a small fraction of samples was used.  When comparing the redundant dictionary to curvelets, the improvement is usually more pronounced when a larger fraction of samples was used.  Notably, for some images, the redundant dictionary reduces the value of the quality metric.  This is more often the case when comparing the redundant dictionary to curvelets and in spite of the fact that curvelet regularization can cause striping artifacts and isolated high values (sometimes described as salt and pepper noise), \cite{wieczorek2015x}.  Thus, even though the metrics are reduced, the images may appear to be better quality to the human eye.  This is the case for the boat reconstruction of Fig. \ref{fig:boatReconsMSBPD}

\section{Discussion and Conclusion}
\label{sec:discussion}

In this work, we show that the low-frequency structure present in both the wavelet and curvelet transforms can be simultaneously exploited to improved the quality when using a redundant dictionary comprised of both transforms.  This exploitation comes in two forms: 1) specifying the size and shape of a fully-sampled center region to satisfy the Nyquist-Shannon sampling theorem, and 2) a Structured BPD problem that determines the details.  We show that this improves the quality over standard compressed sensing reconstructions, and that it improves the quality over a masked version of compressed sensing where the values of the lowest-frequency bins of the wavelet and curvelet transforms are not penalized.  Taking advantage of the fully-sampled center region improved the quality dramatically for the images studied in this manuscript.  The Structured-BPD problem improved the results further still but by a much smaller amount.

The MRI data presented was from three-dimensional data collections that made a hybrid space (two dimensions of frequency and one of location) possible.  For some MRI experiments, the data is collected in such a way that a more general three-dimensional sparsifying transformation is required (e.g. using the cones \cite{gurney2006design} or yarn ball \cite{stobbe2021three} trajectories).  For these applications, the methods of this manuscript can be extended using the three-dimensional discrete curvelet transformation \cite{ying20053d}.

Notably, the low-frequency filter of the the fast discrete curvelet transform of \verb|CurveLab| has a much higher order than that of the DDWT-$4$ wavelet transform.  This suggests that a lower order filter could be used, which would permit a non-negligible amount of aliasing, but where the quality of the reconstruction would be retained when using a fully sampled center region.  This would reduce the computations required to implement the Curvelet transform and its adjoint, which would reduce the time required to attain a reconstruction.

FISTA was used to solve the optimization problems for this manuscript.  However, after exploiting the low-frequency structure, the optimization variable becomes much more sparse.  It may be sparse enough that an Orthogonal Matching Pursuit (OMP) algorithm could be used to solve the problem well.  Candidate algorithms include Compressive Sampling Matching Pursuit (COSAMP) \cite{needell2009cosamp} and Stagewise Orthogonal Matching Pursuit  OMP \cite{donoho2012sparse}.  These are greedy algorithms, so the result may be attained with less time than when using FISTA.


Finally, additional gains in reconstruction quality may be had by either replacing the curvelet sparsifier with wave-atoms \cite{demanet2007curvelets,demanet2007wave} or hexagonal wavelets \cite{zhu2019constructing}, or by augmenting the redundant dictionary with these sets.  We expect these gains to be small, and so their usage would depend on the computational power available and the importance of fine details.

We leave all of these prospects as future work.

\section{Acknowledgments}
ND would like to thank the American Heart Association and the Quantitative Biosciences Institute at UCSF as funding sources for this work.  ND has been supported by a Postdoctoral Fellowship of the American Heart Association.  ND and PL have been supported by the National Institute of Health’s Grant number NIH R01 HL136965.

The authors would like to thank Daniel O'Connor for the many useful discussions on optimization and frames.


\end{document}